\documentclass[a4paper,12pt,reqno,nofootinbib]{revtex4}
\usepackage[centertags]{amsmath}
\usepackage{amsfonts}
\usepackage{amssymb}
\usepackage{amsthm}
\usepackage{newlfont}
\usepackage{stmaryrd}
\usepackage{mathrsfs}
\usepackage{mathtools}
\usepackage{euscript}
\usepackage{graphicx}
\usepackage{enumerate}
\usepackage{todonotes}
\usepackage{color}
\usepackage{floatrow}
\usepackage{caption}

\usepackage{graphicx}

\usepackage{tikz}
\usepackage{pgf}
\usetikzlibrary{positioning,fit,calc}
\usetikzlibrary{arrows,automata}
\usepackage{wrapfig}
\usepackage{subfigure}
\usepackage{amscd}
\usepackage{hyperref}

\theoremstyle{plain}

\theoremstyle{definition}

\theoremstyle{remark}

\newcommand{\opunit}{\text{1}\kern-0.22em\text{l}}

\DeclareMathAlphabet{\mathpzc}{OT1}{pzc}{m}{it}

\newcommand{\id}{\textrm{d}}

\begin{document}

\title{From dynamical systems to statistical mechanics:\\
the case of the fluctuation theorem}

\author{Christian Maes}
\affiliation{Instituut voor Theoretische Fysica, KU Leuven}

\begin{abstract} This viewpoint relates to an article by Jorge Kurchan (1998 J. Phys. A: Math. Gen. {\bf 31}, 3719)
as part of a series of commentaries celebrating the most influential
papers published in the J. Phys. series, which is celebrating its 50th anniversary.
\end{abstract}

\maketitle

\section{Dynamical and statistical explanation}
A recurrent debate in the foundations of statistical mechanics is on the relative importance of dynamical versus statistical arguments.  Even in the work of a single person like Ludwig Boltzmann, at one moment the dynamical view and later the statistical view dominated \cite{stan}.  While a pioneer of fluctuation theory as foundational for thermodynamics, Boltzmann was also the inventor of ergodicity \cite{gal}.  In 1868 he used the ergodic hypothesis to prove equipartition of energy in the kinetic theory of gases.  Maxwell in a letter to Tait in 1873 criticized Clausius and Boltzmann when they aimed at reducing the second law of thermodynamics to a theorem in dynamics, ``as if any pure dynamical statement would submit to such an indignity''\footnote{Maxwell himself had been exposed to the use of probability arguments (mostly in astronomy) ever since Herschel wrote an essay on the {\it Theory of Probabilities} by Adolph Quetelet.}.  On the other hand Boltzmann in 1896 truly hit the meaning of the Maxwell distribution when he commented that ``[the Maxwell distribution]  is in no way a special singular distribution which is to be contrasted to infinitely many more non-Maxwellian distributions; rather it is characterized by the fact that by far the largest number of possible velocity distributions have the characteristic properties of the Maxwell distribution...''  And on the cover of his book  on Lectures in Gas Theory II (1896) Boltzmann cited Gibbs, ``In other words the impossibility of an uncompensated decrease in entropy seems to be reduced to an improbability.''  Thirty years earlier he had claimed to obtain a completely general theorem from mechanics that would prove the second law.\\

Modern theory of dynamical systems started with Henri Poincar\'e, and with other mathematical physicists after him, such as Birkhoff, Hopf, von Neumann,... in the 1920-30's.  Zermelo used Poincar\'e's recurrence theorem (1890) ``against'' Boltzmann's statistical explanation of the second law\footnote{In this anniversary year for Marian von Smoluchowski, it be remembered that as a pioneer of statistical considerations in physics he explained that ``a process appears irreversible if the initial state is characterized by a long
average time of recurrence compared to the times during which the system
is under observation.'' in Smoluchowski, Physik Z. {\bf 17}, 557 (1916).}. In {\it Thoughts on kinetic gas theory} (1906) Poincar\'e discussed the Liouville equation and how (at least a certain form of) entropy must remain constant for isolated Hamiltonian systems\footnote{Of course, the discussion on the origin of dissipation was even much older, {\it e.g.} appearing in the work of d'Alembert (1752) proving that {\it birds cannot fly} (the so called d'Alembert paradox) from the purely mechanical point of view.}.  While statistical ideas were there originally to launch a qualitative study of dynamical systems, {\it e.g.} to understand their typical long time behavior, their invariant measures,  {\it etc.}..., starting from the 1940's people were again rethinking the equilibrium ensembles in terms of ergodic behavior\footnote{A typical example is ``Mathematical Foundations of Statistical Mechanics'' (1949)  of Khinchin.  As Jack Schwartz is remarking ``... the delicious ingenuity of the Birkhoff ergodic theorem has created the general impression that it must play a central role in the foundations of statistical mechanics.'' (in: ``The Pernicious Influence of Mathematics on Science,'' 1962). }.  For the study of steady nonequilibria attention concentrated on entropy production and time-irreversibility.  The dynamical artillery was eventually related  there to notions as  Kolmogorov-Sinai entropy (1958-59) and Policott-Ruelle resonances (1985-86).  The idea was added that there is a breaking of time-reversal invariance because of a different behavior along the stable and unstable manifolds of smooth (Axiom A) dynamical systems.  For example, Sinai-Ruelle-Bowen measures (SRB-states, natural nonequilibrium steady states as they are  sometimes called) are smooth along unstable directions and fractal in the stable direction, which was conjectured to play a grand role in the understanding of production of entropy.  In the 1990's these ideas appeared as foundational for nonequilibrium statistical mechanics \cite{dorf,gasp,gal,hoo,eva}.  There was the feeling that the important aspects of diffusion, transport and dissipation were actually hidden in the Lyapunov spectrum of the underlying dynamical system or in the nature of SRB-states.  And an extension of the ergodic hypothesis was formulated, that ``quite generally a system exhibiting chaotic motions does so in a maximal form so that it can be supposed to be a transitive hyperbolic system.''\cite{schol,chao}.  The idea of that ``chaotic hypothesis'' was then that macroscopic systems, if not under an integrable dynamical system, could be regarded as transitive hyperbolic ``Anosov'' dynamical systems.  We will refer to those also below without further explanations; the reader can imagine smooth deterministic chaotic maps. 

\section{Fluctuation theorem in dynamical systems}
As mentioned before, statistical or probabilistic arguments have always been at the heart of a qualitative understanding of dynamical systems and ergodic theory.  The theory of large deviations for dynamical systems (where the large parameter is time) can be seen as a fluctuation theory around the law of large times (ergodic theorem) \cite{ld1,ld2,ld4,ld3}. The fluctuation theorem in the theory of smooth dynamical systems uses it exactly for the variable phase space contraction rate in reversible dissipative Anosov systems.  In fact, there are various possible versions \cite{2thms,gc,ecm,ja1,ja5}, but one that requires a nontrivial limit procedure is the steady state version, also known as the Gallavotti-Cohen fluctuation theorem \cite{gc,rue,ld3}.\\

In 1993 Evans, Cohen and Morriss
discovered\footnote{Attempting the history of the fluctuation theorems and the problem of attribution is much like entering a minefield, many have felt.  This is no place for such a discussion.} a symmetry in the
fluctuations of the phase space contraction of a thermostated dynamics \cite{ecm}. In the 1980's numerical algorithms (molecular dynamics simulations) for isokinetic and isothermal motion had been developed using deterministic thermostated dynamics \cite{hoo,eva,sar}, {\it e.g.} attempting to simulate the canonical ensemble.  These codes could also be used for driven systems, and it was easily seen (at least in many cases) that the phase space contraction\footnote{For dynamical systems change in Shannon entropy coincides with the phase space contraction rate \cite{andr}.  That change in Shannon entropy for smooth dynamical systems could be maintained in the SRB-state because of the presence of ``fractal directions'' (or non-absolutely continuous pieces) in contrast with the invariance of that Shannon entropy under the Liouville equation.  In that way phase space contraction was naturally associated to entropy production rate.} could in fact be identified with the entropy production \cite{ecm,wag,andr}.  There was no heuristic derivation of these thermostated (non)equilibrium dynamics but their success in simulation was much appreciated and trusted.  Gallavotti and Cohen went on to
prove a fluctuation symmetry for the steady-state distribution of the
time-averages of the phase space contraction rate \cite{gc}.\\ There is a reversible smooth dynamical system $x\mapsto \varphi(x)$ on the phase space $\Gamma$ (a compact
and connected manifold). The transformation $\varphi$ is a diffeomorphism of $\Gamma$ and it is reversible in the sense that there is another diffeomorphism $\pi$ with $\pi^2=1$ for which $\pi\varphi\pi =\varphi^{-1}$.  One assumes sufficient chaoticity in the sense of having a uniform hyperbolicity (and dealing with a transitive Anosov system). There is a unique SRB-state with expectations
\begin{equation}
\langle G \rangle = \lim_{\tau\rightarrow\infty} \frac 1{\tau}\,\sum_{t=0}^\tau G(\varphi_tx)
\end{equation}
meaning time-averages for almost every randomly chosen initial point $x\in \Gamma$.
The logarithm of the Jacobian determinant $D$ corresponding to $\varphi$, $J = −\log D$, is the phase space contraction rate.  One proves (sometimes assumes)
dissipativity, $\langle J\rangle > 0$.  The Gallavotti-Cohen fluctuation theorem is about the fluctuations of 
\begin{equation}
w_\tau(x) = \frac 1{\langle J\rangle\tau}\,\sum_{t=0}^\tau J(\varphi_tx)
\end{equation}
for large time $\tau$.  The theorem states that $w_\tau(x)$ has a distribution $P_\tau(w)$ with
respect to the SRB-state such that
\begin{equation}\label{gcs}
\lim_{\tau\rightarrow\infty}\frac1{\langle J\rangle\tau w }\log \frac{P_\tau[w]}{P_\tau[-w]} = 1
\end{equation}
The stated property is then called the Gallavotti-Cohen symmetry.
We refer to \cite{gc,rue,ld3} for more precise statements.  What interests us here more is that the theorem became famous for nonequilibrium statistical mechanics \cite{rue,gc}. There remained the idea that deterministic chaos was very important for the origin of the second law, and the fluctuation theorem was thought to add a correction to that law; see the title of \cite{ecm}. 
The fluctuation theorem itself was general and non-perturbative but a further reason for the appreciation of the fluctuation theorem indeed came from the derivation of the  Onsager reciprocity and Green-Kubo relations in linear response around equilibrium \cite{onsa}.

\section{Fluctuation theorem in statistical mechanics}
The main observation of Jorge Kurchan in the introduction to his paper \cite{kurch}, is that surely stochastic dynamics are sufficiently chaotic to satisfy the fluctuation symmetry of Gallavotti and Cohen.   In Kurchan's view the ``chaotic hypothesis'' is nothing but demanding stochastic stability, i.e. continuity of expectations with respect to  the addition of some noise.  So he went on testing that symmetry for Langevin dynamics, driven and with a clear meaning of the physical entropy production.  It was the paper that started the broader community of statistical physicists to look more closely for non-pertubative relations in nonequilibrium theory.  The paper was followed by \cite{ls,gibbsian,2000} where the relation with time-reversal was re-enforced and emphasized.\\

Kurchan treats both underdamped and overdamped dynamics.  He emphasizes the importance of boundedness and finiteness, which indeed is an issue for the validity of the asymptotic fluctuation symmetry.  Diffusions on unbounded domains or with unbounded speeds provide a good test case and example.  Towards the end of the paper he also discusses the relation with the fluctuation--dissipation theorem and possible nonlinear extensions. \\

A simple example in \cite{kurch} is taking an underdamped dynamics with damping $\gamma>0$,
\begin{equation}
m\ddot{x} +\gamma \dot{x} + \partial_xU(x) - f(x) = \sqrt{2\gamma T}\,\xi
\end{equation}
where $\xi$ is standard white noise. The force $f$ does not come from a potential (we imagine higher dimensions, multiple particles or nontrivial topology of the domain to make that possible but take a one-dimensional notation for just one particle with mass $m$).  The entropy flux in the thermal environment is $S(\omega) = \int_0^t f(x_s)\cdot v(s)\id s/T$, for a trajectory $\omega=((x_s,v_s), s\in [0,t])$.  Kurchan assumes that there is a non-zero average $\langle S\rangle = \sigma\,t$ in the steady state,  and considers the distribution $\Pi_t$ of $p = S/(\sigma \,t)$.  What he shows is that 
\begin{equation}
\lim_t \frac 1{t} \,\log \Pi_t \rightarrow -\zeta(p),\qquad \zeta(p) -\zeta(-p) =
\sigma p
\end{equation}
where the relation in the right-hand is the Gallavotti-Cohen fluctuation symmetry \eqref{gcs}.\\

By now the fluctuation theorem has been extensively studied, and was the start of many new developments for nonequilibrium statistical mechanics. It is now understood that these fluctuation symmetries, and other relations such as the one of Jarzynski all derive from the same premise and logic which is now summarized as the statement of local detailed balance \cite{gibbsian,ldb3,crooks,poincare}: that for a system in sufficiently weak contact with spatio-temporally sufficiently separated equilibrium reservoirs, the ratio
of probabilities of a system trajectory $\omega$ with respect to its time-reversal $\theta\omega$
satisfies
\begin{equation}\label{tr}
\log \frac{\text{Prob}[\omega]}{\text{Prob}[\theta\omega]} = S(\omega)/k_B + O(1)
\end{equation}
where $S(\omega)$ is the change of entropy in the environment (collection of all equilibrium reservoirs) corresponding to the trajectory $\omega$. The `correction' $O(1)$ stands for a temporal boundary term related to the start and the end of the trajectory $\omega$ but can in principle be unbounded depending on the nature of the state space.  To go from Eq.\eqref{tr} to a Gallavotti-Cohen symmetry for $S(\omega)/k_B$ requires (of course) (1) sufficient statistical ergodicity (e.g. that on the level of the considered physical coarse graining there is a principle of large deviations) and (2) that the correction $O(1)$ can be controlled.  The last issue leads to possible violations of the Gallavotti-Cohen symmetry or to local versions \cite{local,wang}, exactly in the spirit of what Kurchan was addressing in his work.  On the other hand, concerning nonlinear or nonequilibrium extensions of the fluctuation--dissipation theorem the author of the present view believes that the Gallavotti-Cohen symmetry does not play a prominent role \cite{traf}, except for giving nontrivial constraints. Moreover, in the area of ``active particles'' or when the particles  are in contact with nonequilibrium reservoirs \cite{tim}, local detailed balance and hence the fluctuation symmetry for the entropy flux fails except possibly with effective parameters. Kurchan's paper however opened these avenues by bringing the Gallavotti-Cohen symmetry within the context of more standard dynamical fluctuation models that also belong to the conceptual framework of statistical physics.

\end{document}